# Using Software-Defined Networking for Ransomware Mitigation: the Case of CryptoWall

Krzysztof Cabaj[1] and Wojciech Mazurczyk[2]
[1]Warsaw University of Technology, Institute of Computer Science, Warsaw, Poland
[2]Warsaw University of Technology, Institute of Telecommunications, Warsaw, Poland
email: kcabaj@ii.pw.edu.pl, wmazurczyk@tele.pw.edu.pl

*Abstract* — Currently, different forms of ransomware are increasingly threatening Internet users. Modern ransomware encrypts important user data and it is only possible to recover it once a ransom has been paid. In this paper we show how Software-Defined Networking (SDN) can be utilized to improve ransomware mitigation. In more detail, we analyze the behavior of popular ransomware – CryptoWall – and, based on this knowledge, we propose two real-time mitigation methods. Then we designed the SDN-based system, implemented using OpenFlow, which facilitates a timely reaction to this threat, and is a crucial factor in the case of crypto ransomware. What is important is that such a design does not significantly affect overall network performance. Experimental results confirm that the proposed approach is feasible and efficient.

**INTRODUCTION**

Software-defined networking (SDN) is one of the most interesting of the emerging networking paradigms that can overcome the limitations of the current network infrastructures [1]. Thanks to the decoupling of the control and data planes in SDN, the underlying network infrastructure is abstracted from the applications, and therefore the network can be managed in a logically centralized way. Openflow has become the *de facto* standard that implements the SDN paradigm. It is a protocol that allows networking devices, such as switches and routers, which rely on internal flow tables, to be managed by an external controller. All the packets processed by the networking device are compared against its flow table (and the flow entries it consists of). Then, depending on the result, they follow the action for the matching flow entry or are forwarded to the controller if there is no match. Thanks to such a design, the traffic control rules can be applied in a real-time manner.

The introduction of a SDN and the capabilities of this powerful paradigm offer a unique opportunity to provide network security in a more efficient and flexible manner. Consider the following example: when it is discovered that a host is performing malicious activity, then the SDN controller can immediately update the control rules of networking devices. As a result, the host's traffic can be blocked almost instantly, or the device itself can be disconnected from the network. From this perspective, it is not surprising that SDN is currently drawing the attention of cybersecurity researchers and professionals [1]. The first work that proposed a general SDN-based anomaly detection system was put forward by Mehdi et al. in 2011 [2]. Further, others utilized SDN to detect network attacks [3], malicious software behavior on mobile devices [4], or to monitor dynamic cloud networks [5].

In this paper, we present a dedicated SDN-based system to mitigate ransomware threats. Our research concerning various state-of-the-art ransomware families shows that the current trend is to utilize asymmetric cryptography. On the one hand, this prevents a decryption key from residing on a victim's machine, thereby making the threat harder to beat. However, on the other hand, the successful disruption of the communication with the attacker's server can prevent the encryption process. Taking the above into account, we introduce SDN applications that block communication between the victim and the attacker. The proposed system is able to identify suspicious activities through network traffic monitoring and then block infected hosts by applying control rules in a real-time manner and by affecting how network devices handle traffic. To summarize, the main contributions of this paper are:
- The utilization of SDN architecture to mitigate crypto ransomware threats – this enables the development of more specific and effective countermeasures.
- The design of two SDN-based mitigation algorithms that are able to perform efficiently in real-time. These methods rely on our findings from behavioral analyses of one of the most popular ransomwares – CryptoWall.
- The development and evaluation of proof-of-concept implementations of the proposed mitigation system.

The rest of this paper is organized as follows. First, we present ransomware basics, including our findings from behavioral analyses conducted on CryptoWall. Second, we show how popular software and hardware-based security solutions may fail to efficiently counter this threat. Then, we design, develop, and evaluate two SDN-based applications that allow the countering of ransomware in an efficient and timely manner. Finally, we conclude the paper.

**RANSOMWARE BASICS**

Ransomware is malicious software that is designed for direct revenue generation. It is currently an emerging threat for individual users as well as companies and institutions. Symantec reported a 113% increase in ransomware attacks in 2014, and an astounding 4,000% rise in crypto ransomware incidents [6]. In the report "2015 Internet Organized Crime Threat Assessment (IOCTA)," Europol considers ransomware as one of the key threats to the Internet and its users [7]. Additionally, McAfee predicts that in 2016 ransomware will remain a major and rapidly growing threat. Moreover, in May 2015, a ransomware-construction kit called TOX was discovered on the dark web by McAfee Labs [8]. TOX adopted a "ransomware-as-a-service" business model. It allows even inexperienced cybercriminals to create their own customized malware and, using the TOX website (residing on the TOR network), to manage infections and profits. In return for this "service," TOX collects 20% of every ransom paid. Thus, we can speculate that soon there will be even more ransomware infections.

In general, the functioning of typical ransomware is as follows. First, a user machine is infected using various attack vectors, e.g., by drive-by-download, malvertisement, phising, spam, or different forms of social engineering, etc. Then, depending on the type of ransomware, either the victim's machine or the critical data it stores are "locked" until a payment is issued. Moreover, modern versions of this malicious software are able to encrypt all accessible drives, including personal cloud storage services, such as Dropbox, and shared network drives. As a result, it is possible that multiple systems can be compromised by a single infection. The value of an average ransom falls between US$300 and $700, and the favored payment currency is bitcoins [9]. However, there is no guarantee that the user's files/machine will be unlocked afterwards [10].

Typically, modern ransomware can be classified in two main groups: locker and crypto. *Locker ransomware* denies user access to an infected machine. However, it must be noted that in most cases the underlying system and files are left untouched. This means that the malware could potentially be removed without a negative impact on the machine and the stored data. As a consequence, locker ransomware is less effective in achieving its goal compared to the more destructive crypto ransomware. *Crypto ransomware* is a data locker that prevents the user from accessing her/his files or data. The majority of this type of malware relies on the utilization of some form of encryption. After successful infection, typical crypto ransomware covertly searches for and encrypts the files that it deems most valuable (e.g., documents, pictures, and videos, etc.). Clearly, such files are useless until a ransom is paid and the decryption key is obtained. When the encryption process is finished, the victim is presented with an extortion message. The infected machine continues to work normally, as the ransomware does not encrypt any critical system files, so the user is still able to make the payment. Currently, the three most active examples of this kind of malware are CryptoWall, TeslaCrypt (a.k.a. AlphaCrypt), and Locky.

Notably, at first individual users were the main target, but recently a shift towards attacking companies and institutions is observable. This is not surprising, since company desktops and servers are more likely to contain sensitive or critical data, e.g., customer databases, business plans, source code, tax compliance documents, or even webpages. The more valuable the data, the higher the potential ransom, and the greater the chance that it will be paid. This makes companies and institutions highly desired targets. The most famous "success stories" of ransomware infection include law enforcement agencies (Melrose, Tewksbury, and Midlothian Police Departments in the US in 2015-2016) or hospitals (Hollywood Presbyterian Medical Center in the US, Ottawa Hospital in Canada, Lukas Hospital in Neuss, and the Klinikum Arnsberg Hospital in Germany – all in 2016). Unfortunately, in all of the mentioned examples, the victims finally paid the ransom and thus encouraged the cybercriminals to attack and infect more targets.

**CRYPTO RANSOMWARE BASICS: THE CASE OF CRYPTOWALL**

The first cases of modern crypto ransomware appeared around 2005, but initial samples where typically flawed due to the use of custom encryption techniques that could be easily beaten (e.g., the first versions of Trojan.Gpcoder). It should be noted that back then the malware developers typically incorporated symmetric encryption, i.e., the same key was used for both encryption and decryption processes and the key was generated (and often stored) on the victim's machine (Fig. 1, left). Following generation, the key was sent to the Command & Control (C&C) server controlled by the attacker. Communication was intentionally routed through a chain of proxy servers (typically hacked legitimate servers) to conceal the true location of the C&C. Over time, cybercriminals learnt from their mistakes and, by constantly improving the design and the code, they finally turned to asymmetric key cryptography (Fig. 1, right). In this case, a pair of matching cryptographic keys is generated on a C&C server and only the public key is sent to the infected machine. This key is used to encrypt the selected files and the private key never leaves the C&C. This means that, if correctly implemented, asymmetric crypto ransomware is (practically) impossible to break. The most prominent ransomware, and one of the first to introduce asymmetric key cryptography, is CryptoWall 3.0, which was discovered at the beginning of 2015.

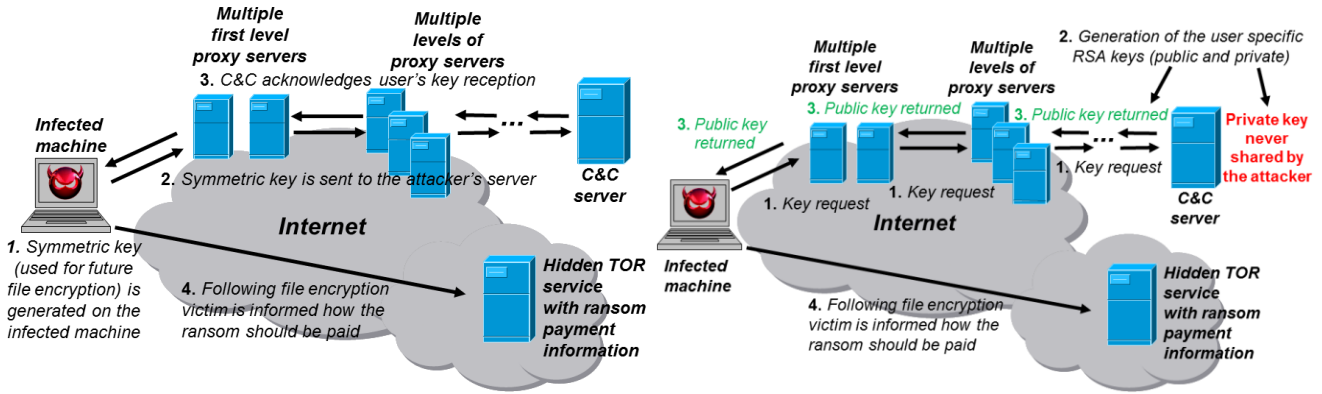

Fig. 1 Symmetric (left) and asymmetric (right) crypto ransomware.

*CryptoWall communications*

From the network traffic perspective, CryptoWall uses domain names instead of direct IP addresses. Due to that, it requires a DNS service to function properly. Analysis of the traffic from infected machines revealed that the first action performed by the CryptoWall 3.0 was learning the victim's IP address using a publicly available service (e.g., ip-addr.es, myexternalip.com, or curlmyip.com). CryptoWall communication utilizes HTTP POST messages directed to the scripts uploaded onto the hacked web servers (proxy servers). This communication is encrypted using the RC4 algorithm and the key is incorporated into the HTTP request. When decrypted, a simple text protocol is revealed (Fig. 2, left). During the first data exchange, the malware reports its unique identifier and the victim's IP address to the C&C, which acknowledges the received information. In the second exchange, the response contains a TOR address of the ransom webpage, the victim's personal code, and an RSA 2048-bit public key that is used for encrypting the data. As mentioned, the public and private keys are generated outside of the infected machine and the proxy. During the third data exchange a PNG image containing instructions for the victim, which will be displayed later, is provided. If the connection is successful, the infected victim acknowledges reception of all the data. Then the communication is suspended due to the ongoing file encryption process. Later, the final, fifth, data exchange contains a number of encrypted files that are presented to the victim with decryption instructions.

Note that there are only slight differences in the protocol utilized by the CryptoWall 3.0 and 4.0 versions (Fig. 2, right). Apart from introducing some data randomization to make its detection harder, the main communication principles remained the same. Moreover, the communication was simplified and, in contrast to the previous version, contains only three message exchanges.

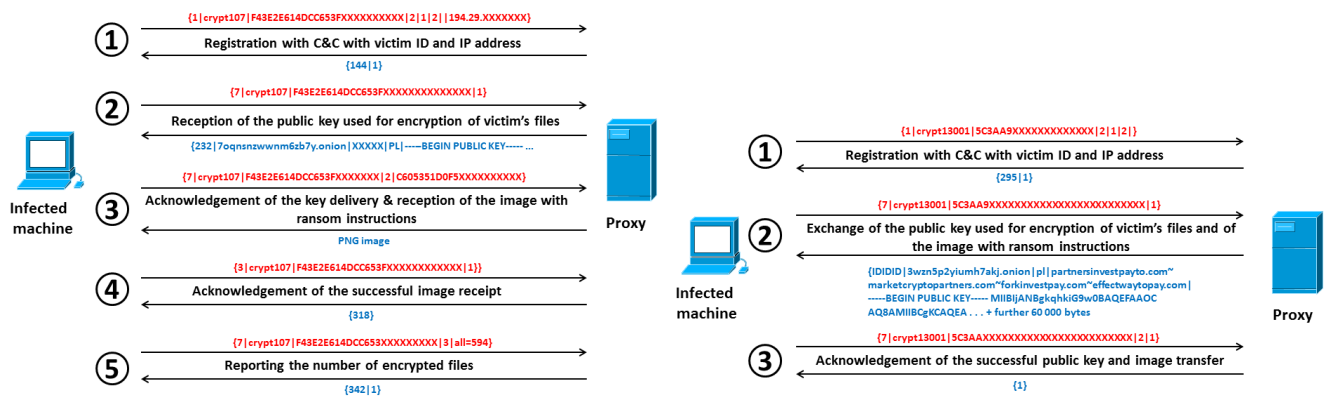

Fig. 2 Communication of CryptoWall 3.0 (left) and 4.0 (right).

*CryptoWall behavioral analysis: main findings*

From the beginning of 2015, we have been analyzing two versions of CryptoWall (3.0 and 4.0) in our customized lab-based malware analysis environment [11]. During the performed investigations, 359 distinct samples were analyzed. They were executed in a controlled environment more than 3,700 times, and more than 5 GB of traffic traces containing connections to proxy servers were captured.

It turned out that when we consider the relationships between CryptoWall domains and IP addresses, typically, a single domain is related to a single IP address (Fig. 3, a, left). However, more complicated structures were also discovered (Fig. 3, a, middle and right). Moreover, to ensure high availability, each CryptoWall sample uses a list of URLs for communication with the C&C server. We observed that the same list of URLs can be used by many distinct samples. As mentioned before, we refer to these lists as "proxy server lists" and we utilize them to cluster all the samples. It is worth noting that the initially observed samples utilized separate proxy server lists. However, some domains and/or URLs were "recycled" later between several lists (Fig. 3, b).

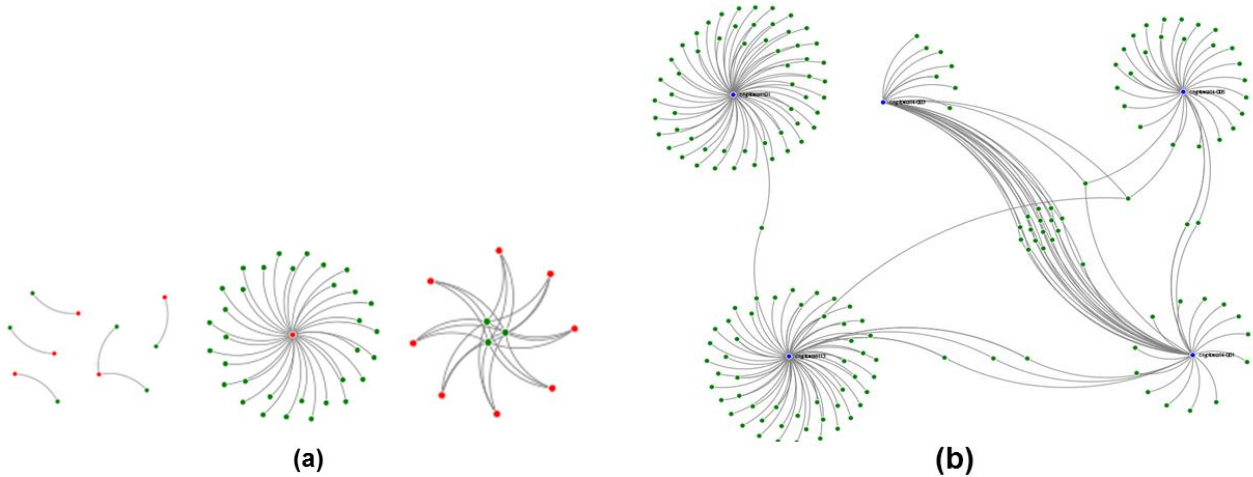

Fig. 3 Relationships between (a) domains (green) and IP addresses (red) and (b) proxy server lists (blue) and domains (green).

Within the 359 analyzed samples we identified 59 distinct proxy server lists. It turned out that the average proxy list contained information for about 40 servers (the largest included 70 servers). The host, when first infected, polls the servers from the list sequentially to find an active one that will serve as a relay to the C&C. It was also revealed that, in the worst case scenario, from the security perspective, i.e., when the malware contacts a responsive proxy server at the first attempt, the time needed for downloading the public encryption key is between 3.76 to 27.38 s (average 9.28 s; median 6.36 s). Moreover, the analyzed data contained 2,038 distinct URLs, which used 1,945 distinct domains. Around 1,700 domains were still successfully resolved in February 2016 – and we detected 1,535 related IP addresses. It is noteworthy that some of the proxy servers were not as short-lived as is commonly believed. One of the longest-lived proxies we encountered was responding with the CryptoWall public key for more than 11 weeks. Note that due to space limitation we share only some of our findings (please see [11] for more detailed analysis).

**MITGATION OF RANSOMWARE THREATS USING CURRENT SECURITY SOLUTIONS: PERFORMANCE ISSUES**

The malware developers behind ransomware are constantly improving their "products," thereby making it harder to develop effective long-lasting countermeasures. It is also foreseen that with the increasing number of devices connected to the Internet (due to, e.g., the *Internet of Things* paradigm), ransomware will soon spread to new device categories [9]. It is also worth noting that, currently, besides desktops and servers, smartphones are being increasingly targeted.

With the growing number of devices plugged into the Internet, the volume of exchanged network traffic has significantly increased. Importantly, the crucial feature of every ransomware countermeasure is the *time* required to detect and react to the discovered malicious activity. In an ideal situation, successful and timely blocking of communication between an infected host and the C&C server can prevent the encryption of the victim's data. Such a requirement poses a serious challenge for the existing security solutions, like firewalls and Intrusion Detection/Prevention Systems (ID/PS), etc., as they typically lack the required level of flexibility and responsiveness. In other words, considering the growing number of devices, the threats and the volume of data, it can be difficult for these security systems to ensure a timely response that will efficiently protect the users.

To show that the currently used security systems do not scale well and do not meet the challenges of modern threats, like ransomware, the following experiments were conducted. First, we explored the performance of a typical software firewall, i.e., the popular *iptables* from Linux. During our experiments we observed an impact from the number of active rules on the overall delay of the traffic passing through the firewall. To evaluate this, we used a one-megabyte test file, which was downloaded from the HTTP server using the *wget* tool, for the different numbers of active rules defined on the firewall (Fig. 4). The experiments were performed on a dedicated machine running Xen hypervisor and all the needed virtual machines: one for downloading the test file, one for the HTTP server and one for the gateway with the firewall. The machine was equipped with an Intel i5-2500K running at 3.3 GHz and 4GB of RAM memory. As it can be observed, when the number of rules did not exceed 10,000, the impact on the network performance could be neglected. However, for higher values there was a major increase in the introduced delays. Our investigation of the CryptoWall ransomware presented in this paper revealed that such a high number of rules may be required if we want to block multiple ransomware families, not to mention other types of malware. Figure 4 also illustrates the relationship between the time needed to apply 1,000 new rules and their total number. The obtained results suggest that the more rules that are active on the firewall, the longer it takes to add additional ones.

The second experiment was similar, only this time we used a hardware firewall, i.e., a Cisco3560 Layer 3 switch. Obviously, a hardware firewall performed better than a software one due to the usage of hardware acceleration with the aid of TCAM memories, which are utilized for the efficient storage and search of rules. However, our experiment revealed that this precious resource can be easily exhausted. The activation of a list with more than 500 rules, which simply denies all IP traffic to/from one hostile IP address, caused full utilization of all TCAM resources. As a result, the switch's operating system decided to process all packets inspected by this list of rules completely with the switch processor, without hardware acceleration. This clearly had a tremendous impact on the forwarding time, as well as on the entire device's performance.

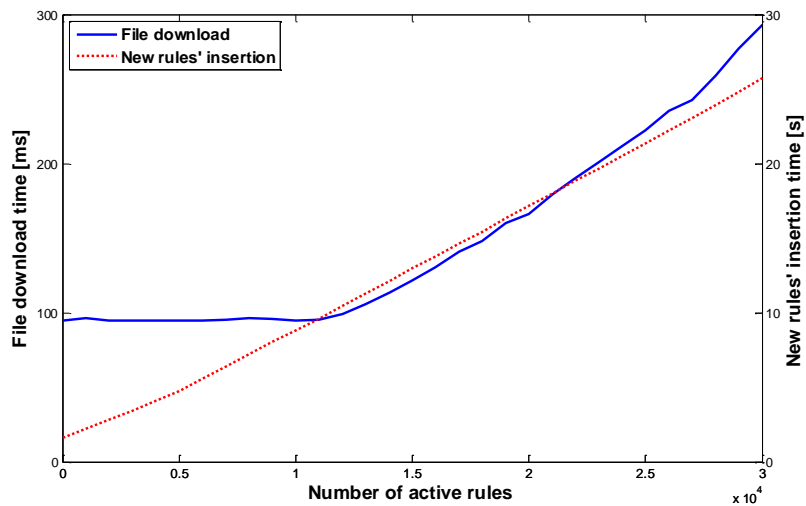

Fig. 4 Performance degradation due to excessive number of active rules and increase in delay due to insertion of 1,000 new rules for popular firewalls.

**PROPOSED MITIGATION METHODS**

Considering the above, our main aim was to evaluate how SDN technology can improve the countering of threats such as ransomware. First of all, it must be noted that the proposed methods were based on CryptoWall findings, but they can also be successfully applied to other types of crypto ransomware (e.g., TeslaCrypt, Locky, etc.). For example, it turns out that the Locky family extends techniques used by CryptoWall to prevent the easy shutdown of the C&C servers by means of simple blacklisting. During the initial phase of the attack, the newest Locky samples utilize few (from two to seven) hardcoded C&C IP address: in most cases these could be shut down rapidly. However, when all hardcoded C&C IPs stop responding, Locky starts querying over a dozen domains generated using DGA (Domain Generation Algorithm). DGA is utilized to periodically generate a number of domain names that can be utilized as rendezvous points with the C&C servers. Our initial findings on Locky indicated that these domains were modified every day. It is worth noting that the principle of the effective mitigation solution stays the same as for the CryptoWall family: without successful connection to the malicious server, the encryption process is not going to be completed, and thus the encryption process would never start.

That is why we proposed, designed, and evaluated two proof-of-concept SDN-based mitigation methods for ransomware, i.e., SDN1 and SDN2. Both rely on dynamic blacklisting of the proxy servers used to relay communication between an infected machine and the C&C server. This well-known method for fighting C&C servers has additional advantages when used for the mitigation of asymmetric-encryption-based ransomware. As mentioned earlier, without access to the C&C server the infected host is not able to retrieve the public key and, as a result, it cannot start the encryption process. However, it must be noted that both of the proposed methods would be efficient only if the ransomware proxy servers were previously identified using, e.g., behavioral analysis of malware samples. For the purpose of this paper we assume that an up-to-date list of proxy servers is available.

The implementation of such a mitigation system can be greatly simplified by developing an SDN application that cooperates with the SDN controller, which is responsible for providing all data needed for the analyses. After the threat is detected, the SDN network infrastructure can be easily reconfigured to block hostile activity and/or to capture the suspicious traffic for further investigation and for recovery of the symmetric encryption key (obviously, this is only possible for symmetric-encryption-based ransomware).

The main functionality of the first application (SDN1) was associated with a simple Layer 2 learning switch. The SDN application forces the switch to forward all DNS traffic to the controller by adding custom crafted flows to the switch during the initialization of the application. As a result, each DNS message is inspected and all responses are sequentially evaluated with the remote database (containing the list of known proxy servers used for ransomware purposes). If the domain name extracted from the DNS message exists in the database, then the response is discarded and it never reaches the infected host, and thus the encryption process will not be performed. In the other case, i.e., when the domain is not listed in the database, no additional actions are performed. When even a single blacklisted proxy server is detected then further communication from this host is not possible, i.e., it is completely blocked and an alert is sent to the system administrator.

The second application (SDN2) was developed to further enhance the performance of the first application. The potential drawback of the SDN1 is that DNS traffic from legitimate and infected hosts is delayed, as for each DNS query the response message is checked with the blacklisted domains database. In the majority of cases there will be no match and the additionally delayed DNS request will be forwarded to the recipient. The SDN2 application solves this problem by introducing custom flows, which forward the DNS query to the intended recipient and only its copy to the controller. This way no additional delay is introduced. While the DNS query is processed, the SDN controller evaluates the extracted domains against the database. If a hostile domain is detected, then the associated IP address is extracted and used to prepare the flow responsible for blocking the traffic between the victim and the C&C server. This flow is then installed in the switch using the OpenFlow protocol. In the other case, i.e., when the domain is not listed in the database, no additional actions are performed.

**EXPERIMENTAL TEST-BED & ANALYSIS**

The testbed used for the experimental evaluation of both implemented SDN applications is presented in Fig. 5. All required software was installed on virtual machines managed by the Xen hypervisor. We used the OpenFlow protocol to control the Open Virtual Switch (OVS) and for forwarding the traffic from all monitored machines. We decided against using OVS as the main switch in the Xen hypervisor to prevent the influence of other management traffic on the experimental traffic. We used the Python-based Pox as a controller, where the functionality of both SDN security applications was implemented.

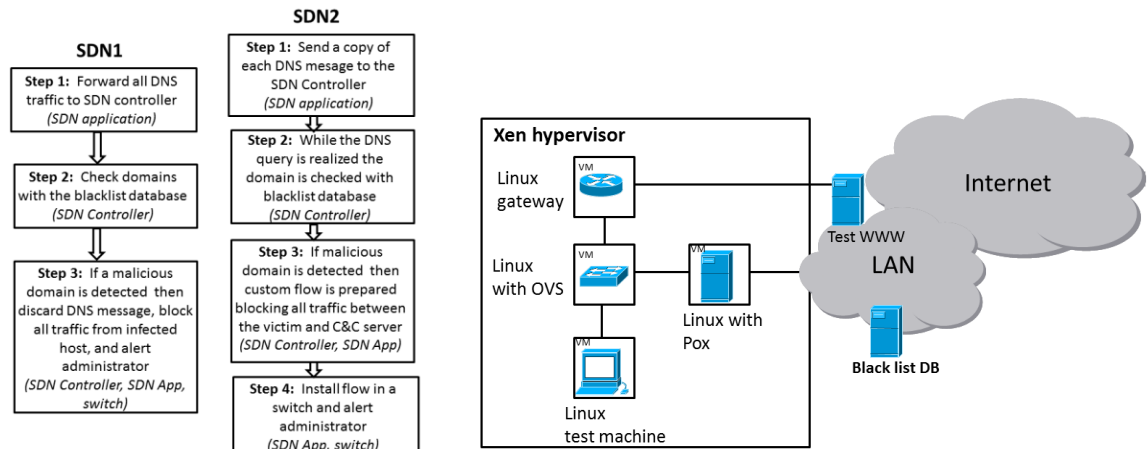

Fig. 5 Description of designed SDN-based applications, SDN1 and SDN2, (left) and the testbed in which they were evaluated (right).

The two proposed proof-of-concept SDN applications were subject to a performance evaluation. As described earlier, the majority of the currently used blacklisting approaches for security devices have a direct impact on network performance. The main reason for this is associated with two facts: *(i)* the need to evaluate each forwarded packet and *(ii)* the usage of rule lists containing all the addresses that should be blocked.

The first advantage of both introduced solutions is associated with the SDN approach to packet forwarding. In currently popular security solutions, all forwarded packets must be evaluated using a defined list of rules. In the SDN, only the first packet from each flow is evaluated, or even a single packet per few flows (the latter approach was adopted in both SDN security applications). The second advantage is related to the fact that rarely within a protected network will all blocked addresses be used simultaneously. SDN allows for the sole implementation of the required rules. As previously mentioned, our CryptoWall analyzes revealed more than 2,100 domains used as proxies for C&C communications. However, not all of these were used by a specific malware sample. Instead of limiting access to all addresses related to these domains, it is possible to block only those that the infected machine tried to contact, which, according to our research, is about 40 domains (on average). As a result, the main requirement is to react in a timely manner to discover if the contacted domain is malicious and to deny all HTTP traffic directed towards it, before the encryption key is received. During the experiments, the time needed to receive a DNS response was measured, as the DNS packets were inspected in the SDN controller before they were sent to the receiving side.

The obtained results concerning the delay introduced to the inspected DNS traffic by both SDN security applications are presented in Fig. 6 (left). To provide reliable results, the first query, which in most cases is directed towards remote DNS servers and could last longer, was omitted. The presented data shows an average value for twenty queries, depending on the number of domains stored in the database. Domains included in the database and used during our experiments represent real malicious domains provided to the Internet community by Bambenek Consulting [12].

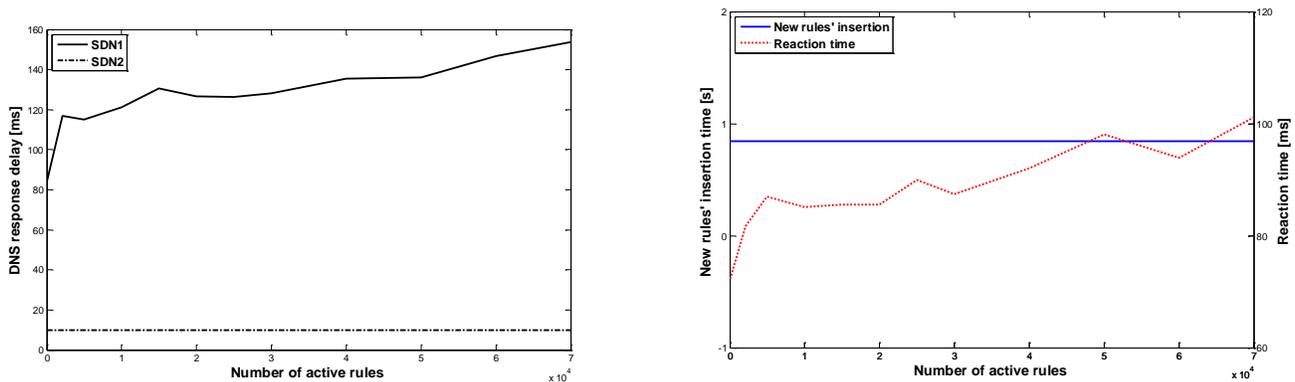

Fig.6 DNS query response delay (left) and the time needed for adding 1,000 new domains for both SDN security applications, and the time needed for the reaction to block a hostile domain with the SDN2 security application (right).

As predicted, the SDN2 application introduced almost no additional delays to the DNS responses and was faster than SDN1. In SDN1 the number of blacklisted domains in the database influenced the DNS query response time. These results cannot be directly compared with the results for the software and hardware-based firewalls presented earlier. However, it should be recalled that for these systems, when the list of rules contained around 30,000 IP addresses, the file download time almost tripled. In contrast, for the SDN solution with 70,000 domains, the increase in response time did not exceed 50%. What should be emphasized, is that the utilization of the domains was more efficient - as our CryptoWall investigation revealed that hostile domains often utilize more than one IP address. Additionally, we observed that some attackers' servers still used the same domains with changed IP addresses, and they still provide malicious content. Another advantage is the utilization of a database, which ensures an almost constant insertion time (Fig. 6, right).

It is worth noting that for SDN2, the time used for verifying the domain is crucial because if it takes too long, then the reaction to the ransomware communication may be too slow. Therefore, the time needed to identify a hostile domain and for the reaction must be investigated. Fig. 6 (right) also presents estimated results, depending on the number of blacklisted domains already stored in the database. Similarly, as for SDN1, the time needed for the reaction was very short, even for tens of thousands of domains it does not exceed 100ms. Note that our CryptoWall analyses revealed that, in the worst case scenario, from the security perspective, the time needed for downloading the public encryption key was between 3.76 to 27.38 s (on average 9.28 s; median 6.36 s). Thus, it is clear that the SDN2 is efficient enough to mitigate ransomware threats, as its reaction time was about 37 times lower than the shortest time that was needed to download the public encryption key. Therefore, we proved that our solution is feasible and efficient for protecting users from the ransomware threat.

## Conclusions

In general, to efficiently fight ransomware, we believe that it is necessary to target and break the business model of the malware developers. Obviously, when the number of infections is lowered, it will significantly decrease the attackers' income and simultaneously increase the operation costs for the attackers' infrastructure. Both of these aspects could be achieved by, for example, rapidly shutting down (i.e., disinfection) the identified proxy servers and by protecting the end users. Obviously, the best protection would be to prevent infection. However, our findings on the CryptoWall ransomware suggest that it is sufficient to be able to disrupt the connection between the victim and the C&C to make the encryption impossible.

That is why in this paper we introduced two SDN-based security applications that can improve the protection against the ransomware. Both rely on an up-to-date database of malicious ransomware proxy servers. The experimental results prove the feasibility and the efficacy of both methods, especially when compared with the performance limitations of the currently most popular security solutions.

Besides the presented approaches, other SDN-based methods can also be envisioned, which could further improve the detection and the countering of ransomware, and which could be combined with the solutions introduced in this paper. These include hybrid approaches, classification-based methods that can detect ransomware based on its well-known network traffic characteristics, and pattern-based methods for automatic detection of the traffic that could be utilized by this type of malware. Both approaches are currently part of our ongoing research on ransomware detection and mitigation.

## References

[1] D. Kreutz, F. V. Ramos, P. Verissimo, C. Rothenberg, S. Azodolmolky, S. Uhlig. "Software-Defined Networking: A Comprehensive Survey", *Proc. of the IEEE*, 103(1):14-76, January 2015

[2] S. A. Mehdi, J. Khalid, S. A. Khayam, "Revisiting Traffic Anomaly Detection Using Software Defined Networking", *Proc. of the 14th International conference on Recent Advances in Intrusion Detection (RAID 2011)*, pp. 161-180, 2011

[3] A. Zaalouk, R. Khondoker, R. Marx, K. Bayarou, "OrchSec: An Orchestrator-Based Architecture For Enhancing Network-Security Using Network Monitoring and SDN Control Functions", *In Proc. of Network Operations and Management Symposium (NOMS)*, pp. 1-9, 2014

[4] R. Jin, B. Wang, "Malware Detection for Mobile Devices Using Software-Defined Networking", *Proc. of GENI Research and Educational Experiment Workshop (GREE '13)*, pp. 81-88, 2013

[5] S. Shin, G. Gu, "CloudWatcher: Network security monitoring using OpenFlow in dynamic cloud networks (or: How to provide security monitoring as a service in clouds?)", *In Proc. of 20th IEEE International Conference on Network Protocols (ICNP)*, USA, 2012, pp. 1-6.

[6] Symantec, "Internet Security Threat Report", April 2015, URL: https://www4.symantec.com/mktginfo/whitepaper/ISTR/21347932_GA-internet-security-threat-report-volume-20-2015-social_v2.pdf

[7] Europol, "Internet Organised Crime Threat Assessment 2015 (iOCTA)", September 2015, URL: https://www.europol.europa.eu/content/internet-organised-crime-threat-assessment-iocta-2015

[8] McAfee Labs, "Meet 'Tox': Ransomware for the Rest of Us", May 2015, URL: https://blogs.mcafee.com/mcafee-labs/meet-tox-ransomware-for-the-rest-of-us/

[9] K. Savage, P. Coogan, H. Lau, "The evolution of ransomware, Symantec Security Response", August 2015, URL: http://www.symantec.com/content/en/us/enterprise/media/security_response/whitepapers/the-evolution-of-ransomware.pdf

[10] A. Kharraz, W. Robertson, D. Balzarotti, L. Bilge, E. Kirda, "Cutting the gordian knot: A look under the hood of ransomware attacks", *12th Conference on Detection of Intrusions and Malware & Vulnerability Assessment (DIMVA 2015)*, July 9-10, 2015, Milan, Italy

[11] K. Cabaj, P. Gawkowski, K. Grochowski, D. Osojca, "Network activity analysis of CryptoWall ransomware", *Przeglad Elektrotechniczny*, vol. 91, nr 11, 2015, ss. 201-204, URL: http://pe.org.pl/articles/2015/11/48.pdf

[12] Bambenek Consulting Feeds: URL: http://osint.bambenekconsulting.com/feeds

## Biography

**Krzysztof Cabaj** holds M.Sc (2004) and Ph.D. (2009) in computer science from Faculty of Electronics and Information Technology, Warsaw University of Technology (WUT). Assistant Professor at WUT. Instructor of Cisco Academy courses: CCNA, CCNP and NS at International Telecommunication Union Internet Training Centre (ITU-ITC). His research interests include: network security, honeypots and data-mining techniques. He is the author or the co-author of over 30 publications in the field of information security.

**Wojciech Mazurczyk** [M'11, SM'13] received the M.Sc., Ph.D. (Hons.), and D.Sc. (habilitation) degrees in telecommunications from the Warsaw University of Technology (WUT), Warsaw, Poland, in 2004, 2009, and 2014, respectively. He is currently an Associate Professor with the Institute of Telecommunications, WUT. His research interests include network security, bio-inspired cybersecurity and networking and information hiding. Since 2013, he has been an Associate Technical Editor of the IEEE Communications Magazine (IEEE Comsoc).